\documentclass[12pt,preprint]{aastex}
\usepackage{verbatim}
 
\begin{document}
 
 \title{Non-Detection of Polarized, Scattered Light \\
 from the HD 189733b Hot Jupiter}
 
 \author{Sloane J. Wiktorowicz}
 \affil{Planetary Science Department, California Institute of Technology, Pasadena, CA 91125 \\
 Current address: Astronomy Department, University of California, Berkeley, CA 94720}
 \email{sloane@berkeley.edu}
   
 \begin{abstract}
Using the POLISH instrument, I am unable to reproduce the large-amplitude polarimetric observations of Berdyugina et al. (2008) to the $> 99.99\%$ confidence level. I observe no significant polarimetric variability in the HD 189733 system, and the upper limit to variability from the exoplanet is $\Delta P < 7.9 \times 10^{-5}$ with 99\% confidence in the 400 nm to 675 nm wavelength range. Berdyugina et al. (2008) report polarized, scattered light from the atmosphere of the HD 189733b hot Jupiter with an amplitude of two parts in $10^4$. Such a large amplitude is over an order of magnitude larger than expected given a geometric albedo similar to other hot Jupiters. However, my non-detection of polarimetric variability phase-locked to the orbital period of the exoplanet, and the lack of any significant variability, shows that the polarimetric modulation reported by Berdyugina et al. (2008) cannot be due to the exoplanet. \\ 
\end{abstract}


\section{Introduction}

  Radial velocity surveys uncover large populations of exoplanets that allow models of planet formation and migration to be constrained. However, in order to study individual exoplanets, it is important to directly detect both their scattered optical flux and their thermal emission. While the Spitzer Space Telescope has enabled exoplanetary thermal emission to be observed (Deming et al. 2005; Harrington et al. 2006, 2007; Knutson et al. 2007, 2009a, 2009b), scattered flux from only one exoplanet has been conclusively observed (Kalas et al. 2008). This is because the contrast ratio between star and exoplanet is at least an order of magnitude larger in the optical than in the infrared. Thermal emission measurements allow exoplanetary temperature maps to be made (Knutson et al. 2007), which constrain models regarding redistribution of stellar insolation by exoplanetary winds. In addition, detection of infrared emission allows molecules such as water vapor (Tinetti et al. 2007) and methane (Swain et al. 2008) to be identified in exoplanetary atmospheres. Atomic species can be identified in these atmospheres by transmission spectroscopy in the optical, and it can also be used to identify the presence of cloud layers (Charbonneau et al. 2002; Vidal-Madjar et al. 2003; Barman 2007; Pont et al. 2008; Redfield et al. 2008). If the exoplanetary radius can be estimated, detection of light scattered by an exoplanet allows its geometric albedo to be determined. Geometric albedo is a measure of the scattering in the atmosphere of the exoplanet, which gives information about atmospheric cloud structure. \\
  
  However, most thermal emission and transmission spectroscopy measurements of exoplanets are from transiting systems, where the orbital plane is seen edge-on. This is because thermal emission may be identified during secondary eclipse, when the exoplanet is occulted by the star, and transmission spectroscopy requires a primary eclipse by definition. Transits allow baseline stellar emission (both in the optical and infrared) to be subtracted from the combined star/exoplanet signal, which greatly improves the signal to noise ratio for direct detection. Unfortunately, transiting systems only comprise about 10\% of massive, short period exoplanets (so-called ``hot Jupiters"), so high signal to noise observations of the atmospheres of most known exoplanets are not accessible with these techniques. While advanced imaging instruments, such as the Gemini Planet Imager (Macintosh et al. 2006; Graham et al. 2007b), have the potential to observe thermal emission from long-period exoplanets, the majority of known exoplanets orbit too closely to their host stars in order to be accessible to imaging. \\

  In addition to observing scattered and emitted flux from exoplanets, determining accurate exoplanetary masses is needed to characterize individual exoplanets. However, the dominant exoplanet-finding technique, radial velocity, is insensitive to stellar reflex motion in the plane of the sky. Therefore, measured mass $m$ is only a lower limit to the true mass $M$ because $m = M \sin i$. While one may infer the most probable exoplanetary mass by assuming an isotropic distribution of orbital inclination $i$, an observational technique that constrains inclination is desired. \\
  
  Again, transiting systems are a boon to the characterization of individual exoplanets, because the shape of the transit lightcurve is indicative of both orbital inclination and exoplanetary radius. Inclination estimates from transit observations can be coupled with radial velocity data to derive accurate masses. Indeed, the mass of the transiting hot Jupiter HD 209458b has been measured to within five Earth masses (Torres et al. 2008 analysis of Brown et al. 2001; Naef et al. 2004; Butler et al. 2006). The hot Neptune GJ 436b, with an M dwarf host, has an uncertainty of less than one Earth mass (Torres et al. 2008 analysis of Deming et al. 2007; Gillon et al. 2007a, 2007b; Maness et al. 2007). Knowledge of accurate exoplanetary radii enables bulk exoplanetary density to be determined. For example, the relatively large mass with respect to radius of the transiting exoplanet HD 149026b indicates the presence of a large, rocky core, which bolsters support for its formation via core accretion (Sato et al. 2005). In order to determine accurate exoplanetary mass for non-transiting exoplanets, however, new observational techniques are necessary.  \\
   
   \section{Exoplanetary Polarimetry}
  
  Polarimetric observations have the potential to determine orbital inclination, and therefore accurate mass, for exoplanets by isolating their scattered, optical flux. Indeed, polarimetry has been utilized to study dust grains in debris disks because of its sensitivity to scattered light (Meyer et al. 2002; Jensen et al. 2004; Monin et al. 2006; Perrin et al. 2006; Graham et al. 2007a; Beckford et al. 2008). The extra information content offered by polarimetry over photometry suggests that exoplanetary scattered flux can be detected even in non-transiting systems. That is, photometric detection of exoplanetary flux benefits greatly from chopping of exoplanetary flux between in-transit and out of transit observations due to the scalar nature of photometric observations. However, the vector nature of polarimetric observations may enable direct detection of exoplanetary flux even when system intensity is constant. \\
  
\begin{figure}[t]     
 \begin{center}
  \scalebox{0.8}{\includegraphics{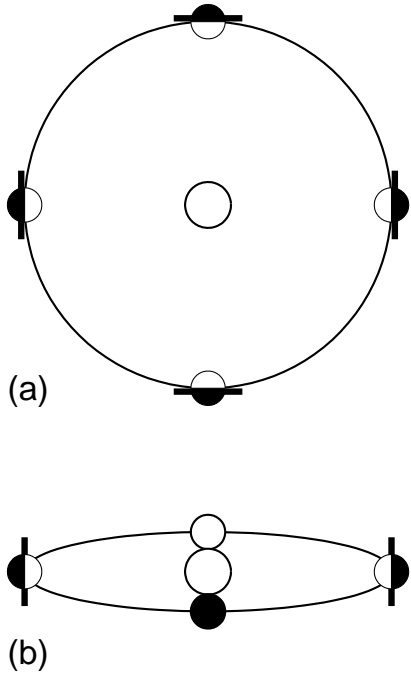}}
  \end{center}
 \caption{Schematic orbital modulation of polarization for an exoplanet with (a) face-on and (b) edge-on geometries. The amount of light scattered by the exoplanet is represented by the white, illuminated portion of the exoplanet, and the degree of polarization is strongest when the exoplanet is near quarter phase. The position angle of net polarization is given by the orientation of the black lines.}   
 \label{1figb}        
\end{figure}

  Polarization of exoplanets arises by scattering of incident starlight by gas molecules, aerosols, and dust grains in the atmosphere of the exoplanet (Seager et al. 2000). For a face-on, circular orbit (Figure \ref{1figb}a), the exoplanet is always seen at quarter phase and always has half its disk illuminated. For a featureless exoplanetary atmosphere, the intensity of light scattered by the exoplanet is constant throughout the orbit and the degree of polarization is also constant. However, the position angle of polarization rotates through $360^\circ$ each orbit, because the scattering plane rotates as the exoplanet progresses in its orbit. \\
  
  In contrast, an edge-on viewing geometry generates large, periodic variability in the degree of polarization because the amount of scattered light is variable (Figure \ref{1figb}b). However, the scattering plane is always nearly coplanar with the orbital plane, so the position angle of net polarization does not vary significantly throughout the orbit. It should be noted that the polarimetric signature of an exoplanetary transit is discussed in Carciofi \& Magalh\~{a}es (2005) and in section 4.2. In general, an exoplanet exhibits variability in the polarization vector that is indicative of orbital inclination, and the models of Seager et al. (2000) and Stam et al. (2004) demonstrate this effect. \\

The simplest discussion of the modulation in system polarization due to an exoplanet exists for a Lambertian phase function and a polarization dependence on scattering angle similar to Rayleigh scattering. This is given by

\begin{equation}
P(\phi) = \epsilon F(\phi) P_0 (\phi).
\label{eqa}
\end{equation}

\noindent Here, $\phi$ is orbital phase pinned to stellar radial velocity phase ($\phi = 0$ represents superior conjunction of the exoplanet), $\epsilon = p\left(R_p / a\right)^2$ is the fraction of stellar flux scattered by an exoplanet with radius $R_p$ and semimajor axis $a$ (assuming a circular orbit), $p$ is its geometric albedo (the fraction of exoplanetary scattered flux at full phase compared to that scattered by a Lambertian disk), $F(\phi)$ is the phase function (the fraction of intercepted flux scattered toward the observer), and $P_0 (\phi)$ is the polarization of that scattered flux. The phase function is given in terms of the phase angle $\alpha$, which is the angle between the host star and observer as seen from the exoplanet. Russell (1916) gives the Lambertian phase function as the following:

\begin{equation}
F(\alpha) = \frac{\sin \alpha + (\pi - \alpha) \cos \alpha}{\pi}.
\end{equation}
\smallskip

\noindent Given $\cos \alpha = \sin i \cos \phi$, $F(\phi)$ can be determined. It is useful to decompose $P(\phi)$ in terms of its normalized Stokes parameters, where $P(\phi) = \sqrt{Q^2(\phi) + U^2(\phi)}$ and

\begin{mathletters}
\begin{equation}
Q'(\phi) = \epsilon F(\phi) (\sin^2 \phi - \cos^2 \phi \cos^2 i)
\label{eqb}
\end{equation}
\begin{equation}
U'(\phi) = \epsilon F(\phi) \sin 2 \phi \cos i
\label{eqc}
\end{equation}
\end{mathletters}

\noindent after Shakhovskoi (1965). The primed Stokes parameters are measured in the orbital frame and $+Q'$ is defined to be in the direction of the orbital angular momentum vector. Converting polarization from the orbital frame to celestial coordinates requires knowledge of the longitude of the ascending node $\Omega$, where $\theta = \Omega + 270^\circ$ (Figure \ref{3figr}):

\begin{figure}[t]     
 \begin{center}
  \scalebox{0.7}{\includegraphics{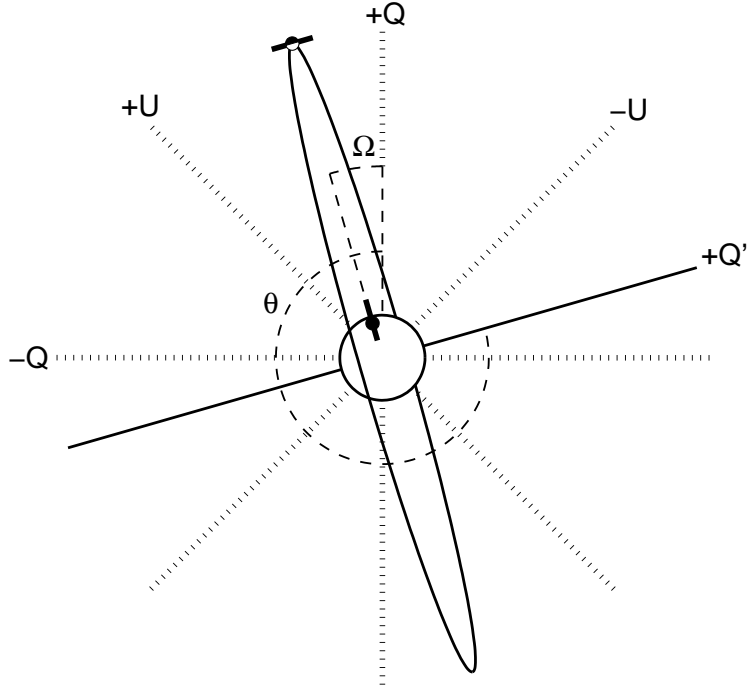}}
  \end{center}
 \caption{Schematic polarization due to an exoplanet at quarter phase and to an equatorial starspot on the stellar limb (section 4.2). Celestial North is in the direction of $+Q$, and East is in the direction of $-Q$. Sizes of the star, exoplanet, and orbit are to scale for the HD 189733 system. The value of $\Omega = 16^\circ$ is taken from Berdyugina et al. (2008).}   
 \label{3figr}        
\end{figure}           

\begin{mathletters}
\begin{equation}
Q(\phi) = Q'(\phi) \cos 2 \theta - U'(\phi) \sin 2 \theta
\end{equation}
\begin{equation}
U(\phi) = Q'(\phi) \sin 2 \theta + U'(\phi) \cos 2 \theta.
\end{equation}
\end{mathletters}

While terrestrial planets are expected to follow $\Delta P \equiv P(\phi)_{\rm max} - P(\phi)_{\rm min}\propto p$ because of their solid scattering surfaces, multiple scattering reduces the polarization of light scattered by gas giant atmospheres with high albedos (Seager et al. 2000; Stam et al. 2004; Lucas et al. 2009). Thus, the form of polarization as a function of orbital phase from Equations 1 through 4 is expected to approximate that of all exoplanets, but the amplitude of the signal will be strongly overestimated for gas giants. \\

From the Monte Carlo scattering simulations of Lucas et al. (2009), the polarization amplitude of a gas giant exoplanet is expected to be

\begin{mathletters}
\begin{equation}
\Delta P_{i = 90^\circ} = (-2.10 p^2 + 2.91 p) \left(\frac{a}{0.05 \, \rm{AU}}\right)^{-2} \left(\frac{R_p}{1.2 \, \rm{R_J}}\right)^{2} \!\!\! \times 10^{-5}
\label{eqd}
\end{equation}
\begin{equation}
\Delta P_{< i >} = 1.43 (\Delta P_{i = 90^\circ}).
\label{eqe}
\end{equation}
\end{mathletters}

\noindent Equation \ref{eqd} is the polarization amplitude for an edge-on orientation, while Equation \ref{eqe} relates to the general case where orbital inclination is unknown. The quadratic fit to $\Delta P$ as a function of $p$ reproduces the $<\!\! P_{MV} \!\!>$ values in Table 6 of Lucas et al. (2009) to within 2\%. \\

 The strongest polarimetric modulation from a gas giant exoplanet will occur for a Lambertian phase function with a polarization dependence on scattering angle similar to Rayleigh scattering. Here, $p = 2/3$ at all wavelengths. Since $(R_p / a)^2 = (3.081 \pm 0.043) \times 10^{-4}$ for HD 189733b (Torres et al. 2008 analysis of Knutson et al. 2007), Equation \ref{eqd} shows that the amplitude of polarimetric modulation due to the exoplanet cannot be larger than $\Delta P \approx 2 \times 10^{-5}$ for scattered light at all wavelengths. Upper limits to geometric albedo from reflected light campaigns exist for the exoplanets around $\tau$ Bo\"{o}, HD 75289, and HD 209458 from photometry (Leigh et al. 2003a, 2003b; Rowe et al. 2008b). Lucas et al. (2009) estimate an upper limit for $\tau$ Bo\"{o} b using PlanetPol, a polarimeter capable of part per million precision on bright stars (Hough et al. 2006). These albedo upper limits, along with the bandpass in which they were determined, are shown in Table \ref{table4}. Assuming a mean albedo upper limit of $p \la 0.22$, the polarimetric modulation from HD 189733b is $\Delta P_{\rm exp} \la 1 \times 10^{-5}$. Since the albedo estimates in Table \ref{table4} range from 400 to 920 nm, this expected upper limit to HD 189733b modulation is for scattered light at all optical wavelengths. \\

\begin{deluxetable}{c c c c c c c}
\tabletypesize{\small}
\tablecaption{Upper Limits to Exoplanetary Albedos}
\tablewidth{0pt}
\tablehead{
\colhead{Target} & \colhead{$p$} & \colhead{Confidence} & \colhead{$\lambda_{\rm min}$} & \colhead{$\lambda_{\rm max}$} & \colhead{Method} & \colhead{Ref} \\
& & & (nm) & (nm) & &}
\startdata
HD 75289 & $< 0.12$ & 99.9\% & 406.5 & 522.0 & Photometry & 1 \\
HD 209458 & $< 0.17$ & $3\sigma$ & 400 & 700 & Photometry & 2 \\
$\tau$ Bo\"{o} & $< 0.37$ & $4\sigma$ & 590 & 920 & Polarimetry & 3 \\
$\tau$ Bo\"{o} & $< 0.39$ & 99.9\% & 407.4 & 649.0 & Photometry & 4 \\
\enddata
\label{table4}
\tablerefs{(1) Leigh et al. (2003b), (2) Rowe et al. (2008b), (3) Lucas et al. (2009), (4) Leigh et al. (2003a).}
\end{deluxetable}

\section{Observations and Data Reduction}
 
 The POLISH instrument (POLarimeter for Inclination Studies of High mass x-ray binaries/Hot jupiters) is a visible light polarimeter commissioned at the Cassegrain focus of the Hale 5-m telescope at Palomar Observatory, California. This instrument is described in Wiktorowicz \& Matthews (2008, hereafter WM08). It utilizes a photoelastic modulator (PEM), a Wollaston prism, lock-in amplifiers, and digital voltmeters to modulate and detect incident, polarized light at 100 kHz. The Wollaston prism feeds a pair of avalanche photodiodes (APDs) or photomultiplier tubes (PMTs), depending on stellar intensity. The bandpass of the instrument is limited by the detectors and atmospheric transmission; the lack of spectral filters increases throughput of the instrument and allows for high precision observations. The APD bandpass ranges from about 400 to 850 nm, while the PMT bandpass ranges from roughly 400 to 675 nm. Atmospheric transmission sets the lower limit for spectral coverage for both detector types. The above components and large telescope aperture contribute to the high signal-to-noise observations with the instrument, where part per million precision is achieved on bright stars (WM08). On-source guiding is accomplished by use of a beamsplitter, which allows about $5\%$ of the flux to be sent to a CCD camera.\\
 
The voltage output from the detectors consists of a roughly sinusoidal waveform with a frequency of 100 kHz and a DC offset. Polarization of the input beam is proportional to the ratio of the AC amplitude (polarized flux) to the DC offset (total flux). Thus, atmospheric effects that operate on timescales longer than 10 $\mu$s are effectively removed. For example, non-birefringent cirrus clouds passing through the telescope beam will reduce both polarized and unpolarized flux equally: the ratio of AC to DC signals will remain constant. Additionally, polarization imparted from such forward scattering by cloud particles will be zero. Therefore, high precision, integrated light, polarimetric observations with POLISH are photon shot noise limited (WM08), eliminating the benefit from a similar, space-based polarimeter. \\

The amplitude of the 100 kHz AC signal is sampled at 8 Hz by the lock-in amplifiers, and the DC offset is sampled at 6 Hz by the voltmeters. The fact that the sampling rates are different is not important because polarization is related to time-averaged AC and DC values. Each on-source measurement consists of one, 30 second integration during which the AC amplitude and DC offset are sampled. AC and DC signals are then sky subtracted by chopping the secondary mirror to a sky field and initiating another 30 second integration. It should be noted that no less than 30 seconds elapse before the next chop throw. Since sky background is low in both the AC amplitude and DC offset, asymmetric, 2:1 source/sky chops are used. That is, observing ``triplets" consist of repeated sequences of source, sky, and source integrations every 90 seconds. \\

The Cassegrain ring is rotated by $45^\circ$ roughly every 10 minutes, which causes a different linear Stokes parameter to be sampled. This ensures that all $\pm Q$ and $\pm U$ Stokes parameters are sampled at least once for each star to minimize systematic effects. Note that each star is observed for about one hour per night. Nightly telescope polarization of $3.0 \times 10^{4}$ is then subtracted. Telescope polarization is the mean of the nightly measurements of the bright ($V \approx 3$), unpolarized star HR 5854 weighted by the DC level of each measurement. DC level is proportional to the number of detected photons, and weighting by this quantity ensures that each photon is weighted equally. This is important in partly cloudy conditions, because observed stellar intensity varies throughout the night. \\

The equatorial mount of the Hale 5-m inhibits traditional telescope polarization measurement, which involves allowing the field to rotate and determining the center of the $(Q,U)$ locus for weakly polarized targets. However, Hough et al. (2006) and Lucas et al. (2009) performed this analysis and report a polarization of four parts per million for HR 5854. Lucas et al. (2009) report that the polarization of this star varies by seven parts per million over one year, but this lies at the $< 3 \sigma$ level of significance. Therefore, I assume that this star is indeed unpolarized, and I assume that it is non-variable over this seven-night observing run. Since telescope polarization is a function of wavelength, it will be different for stars with different spectral types. In addition, I quickly measure telescope polarization to the part per million level with APDs on a bright, unpolarized star, even though the science targets in this work are about five magnitudes fainter and are observed with the different bandpass of the PMTs. Therefore, an error of order one part in $10^5$ or less in the absolute calibration of telescope polarization may be present. However, such an offset is assumed to be non-variable and below the measurement noise for the targets in this work. Systematic effects have been reduced to the part-per-million level for weakly polarized sources and to less than 1\% of the measured polarization for significantly polarized sources (WM08). \\

Nightly mean and run-averaged polarizations for each target are determined by taking the mean of the polarization measurements weighted by their DC levels. Uncertainty in polarization is the square root of the weighted variance of the polarizations from these measurements divided by the square root of the number of measurements. Each polarization measurement is photon shot noise limited, and uncertainty in run-averaged polarization scales with the square root of stellar polarization. The polarization noise floor of the instrument is about two parts per million (WM08). I calibrate absolute polarization measured with POLISH against the catalog of Heiles (2000). However, since my observed position angles of net polarization are generally consistent with those in the literature for strongly polarized targets, I do not calibrate position angle. It should be stressed that the unfiltered polarimetric observations of science targets in this work are taken over the wavelength range of roughly 400 to 675 nm. \\
 
\begin{deluxetable}{c c c c c c c c c}
\tabletypesize{\scriptsize}
\tablecaption{Observed Stars}
\tablewidth{0pt}
\tablehead{
\colhead{Name} & \colhead{Alt. Name} & \colhead{RA} & \colhead{Dec} & \colhead{$P$} & \colhead{$\Theta$} & \colhead{Ref} & \colhead{$V$} & \colhead{Type} \\
& & (J2000) & (J2000) & ($\%$) & ($^\circ$) & & & }
\startdata
HR 5854 & $\alpha$ Ser & $15^{\rm h}44^{\rm m}16\, \fs07$ & $+06\, \fdg25 \arcmin 32\, \farcs 3$ & $0.00020(22)$ & $-$ & 1 & 2.64 & K2IIIb\\
HD 149026\tablenotemark{a} & SAO 65349 & $16^{\rm h}30^{\rm m}29\, \fs62$ & $+38\, \fdg20 \arcmin 50\, \farcs 3$ & ? & ? & $-$ & 8.16 & G0IV \\
HD 175541\tablenotemark{b} & GJ 736 & $18^{\rm h}55^{\rm m}40\, \fs88$ & $+04\, \fdg15 \arcmin 55\, \farcs 2$ & ? & ? & $-$ & 8.03 & G8V \\
Cygnus X-1\tablenotemark{c} & SAO 69181 & $19^{\rm h}58^{\rm m}21\, \fs68$ & $+35\, \fdg12 \arcmin 05\, \farcs 8$ & 4.98(18) & 135.0(1.0) & 2 & 8.95 & O9.7Iab\\
HD 189733\tablenotemark{a} & V452 Vul & $20^{\rm h}00^{\rm m}43\, \fs71$ & $+22\, \fdg42 \arcmin 39\, \farcs 1$ & 0.0212(30) & 99.6(4.1) & 3 & 7.68 & K1.5V \\
HR 8974\tablenotemark{b} & $\gamma$ Cep & $23^{\rm h}39^{\rm m}20\, \fs85$ & $+77\, \fdg37 \arcmin 56\, \farcs 2$ & $0.00047(35)$ & $-$ & 1 & 3.23 & K1IV \\
\enddata
\label{3tablea}
\tablenotetext{a}{Transiting, short period exoplanet host}
\tablenotetext{b}{Non-transiting, long period exoplanet host}
\tablenotetext{c}{High mass X-ray binary}
\tablerefs{(1) WM08, (2) Heiles (2000), (3) B08}
\tablecomments{WM08 polarizations are measured in the wavelength range of about 400 to 850 nm, Heiles (2000) polarization data are quoted for $V$ band, and B08 polarization is measured in $B$ band.}
\end{deluxetable}
 
 The stars observed are listed in Table \ref{3tablea}. Polarimetric data are obtained from the catalogs of Heiles (2000), from Berdyugina et al. (2008, hereafter B08), and from WM08. All other non-polarimetric data are taken from the SIMBAD database. The polarization and position angle values in parentheses represent the standard error of the mean, which assumes nonvariable target polarization. However, Cygnus X-1 is known to be variable with a polarimetric amplitude of order $\Delta P \approx 0.1\%$ (Nolt et al. 1975; Kemp 1980; Dolan \& Tapia 1989). The amplitude of variability of Cygnus X-1 is also variable and due to both stable and stochastic effects (Dolan \& Tapia 1989, 1992). This variability is roughly phase-locked to the orbital period of the binary, and Cygnus X-1 is included in this work as a variable control source. \\
 
 The long period exoplanet hosts HD 175541 and HR 8974 are included as non-variable control sources, as the exoplanets should only be detectable over an entire orbit and at the level of $\Delta P_{\rm exp} \la 2 \times 10^{-8}$ and $\Delta P_{\rm exp} \la 5 \times 10^{-9}$, respectively (Equation \ref{eqe}). The transiting hot Saturn around HD 149026 is expected to generate a polarimetric signal of $\Delta P_{\rm exp} \la 2 \times 10^{-6}$ over its orbit (Equation \ref{eqd}). Orbital parameters and expected polarizations of these systems are given in Table \ref{table3}. These amplitudes assume a geometric albedo upper limit of $p \la 0.22$ (the mean of the upper limits in Table \ref{table4}) and a radius of $R = 1.2$ $R_{\rm J}$ for the non-transiting exoplanets. Since these upper limits to polarimetric modulation range from 400 to 920 nm, the expected upper limits are for scattered light at all optical wavelengths. No polarimetric data for HD 149026 or HD 175541 were found in the literature.  \\

\begin{deluxetable}{c c c c c c c}
\tabletypesize{\small}
\tablecaption{Expected and Observed Variability}
\tablewidth{0pt}
\tablehead{
\colhead{Target} & \colhead{$R$} & \colhead{$a$} & \colhead{$i$} & \colhead{Ref} & \colhead{$\Delta P_{\rm exp}$} & \colhead{$\Delta P_{\rm obs}$} \\
& $(R_{\rm J})$ & (AU) & ($^\circ$) & & & }
\startdata
HR 8974 & $\approx 1.2$ & $2.044(57)$ & ? & 1 & $\la 5 \times 10^{-9}$ & $9.0 \times 10^{-6}$ \\
HD 175541 & $\approx 1.2$ & $1.03(-)$ & ? & 2 & $\la 2 \times 10^{-8}$ & $2.6 \times 10^{-5}$ \\
HD 149026 & $0.654(^{+60}_{-45})$ & $0.04313(^{+65}_{-56})$ & $90(^{+0.0}_{-3.0})$ & 3 & $\la 2 \times 10^{-6}$ & $1.6 \times 10^{-5}$ \\
HD 189733 & $1.138(27)$ & $0.03099(^{+60}_{-63})$ & $85.58(6)$ & 4 & $\la 1 \times 10^{-5}$ & $2.1 \times 10^{-5}$ \\
Cygnus X-1 & $-$ & $0.195(42)$ & $48(7)$ & 5 & $\approx 10^{-3}$ & $5.1 \times 10^{-4}$ \\
\enddata
\label{table3}
\tablerefs{(1) Neuh\"{a}user et al. (2007); (2) Johnson et al. (2007); (3) Torres et al. (2008) analysis of Sato et al. (2005), Butler et al. (2006), Charbonneau et al. (2006, Winn et al. (2008); (4) Torres et al. (2008) analysis of Bouchy et al. (2005), Knutson et al. (2007); (5) Iorio (2008) analysis of Gies et al. (2003), Shaposhnikov \& Titarchuk (2007).}
\tablecomments{Expected variability for all targets is in broadband $BVRI$ light. The observed upper limit to variability of HR 8974 is in the wavelength range of about 400 nm to roughly 850 nm, and upper limits for the rest of the sample are measured between 400 and 675 nm. References are for orbital data.}
\end{deluxetable}
 
\section{Results and Discussion}
\subsection{Unbiased Significance of Variability}

Nightly mean and run-averaged polarizations for the observed targets are listed in Table \ref{table2} after telescope polarization (hereafter ``TP") is subtracted. Note that run-averaged polarization assumes non-variability of the source. This is apparent in the formal uncertainty in polarization for Cygnus X-1, $\sigma_P \approx 4 \times 10^{-5}$, because this source is known to be variable on the order of $\Delta P \approx 0.1\%$. I refer the reader to Table 4 of WM08 for nightly polarization measurements of the unpolarized standard star HR 8974. The standard deviation of mean nightly polarizations is given as $\Delta P_{\rm obs}$ in Table \ref{table3}. These represent measurement precision coupled with intrinsic source variability. \\

\begin{deluxetable}{c c c c c c}
\tabletypesize{\scriptsize}
\tablecaption{Observed Polarization}
\tablewidth{0pt}
\tablehead{
\colhead{UT Date} & \colhead{Target} & \colhead{$Q/I$} & \colhead{$U/I$} & \colhead{$P$} & \colhead{$\Theta$} \\
& & $(\%)$ & $(\%)$ & $(\%)$ & $\left(^\circ\right)$ }
\startdata2008 Jun 08 & TP & $-$0.02645(29) & +0.01637(40) & 0.03111(32) & 74.12(34) \\
2008 Jun 09 & $\cdots$ & $-$0.02211(50) & +0.01370(59) & 0.02601(53) & 74.10(63) \\
2008 Jun 10 & $\cdots$ & $-$0.02630(41) & +0.01624(50) & 0.03091(43) & 74.15(44) \\
2008 Jun 11 & $\cdots$ & $-$0.02625(27) & +0.01634(39) & 0.03092(31) & 74.05(33) \\
2008 Jun 12 & $\cdots$ & $-$0.02597(29) & +0.01622(43) & 0.03062(33) & 74.01(37) \\
2008 Jun 13 & $\cdots$ & $-$0.02578(26) & +0.01576(41) & 0.03022(31) & 74.29(36) \\
2008 Jun 14 & $\cdots$ & $-$0.02599(27) & +0.01657(36) & 0.03082(30) & 73.74(31) \\
Overall & $\cdots$ & $-$0.02578(16) & +0.01605(18) & 0.03037(16) & 74.05(16) \\
\hline
2008 Jun 08 & HD 149026 &   $-$0.0438(61) & +0.0142(43) &   0.0460(59) &  81.0(2.8) \\
2008 Jun 09 & $\cdots$ &   $-$0.0445(19) &  +0.0053(35) &   0.0448(19) &  86.6(2.2) \\
2008 Jun 10 & $\cdots$ &   $-$0.0425(21) &  +0.0084(31) &   0.0433(21) &  84.4(2.0) \\
2008 Jun 11 & $\cdots$ &   $-$0.0402(20) &  +0.0045(32) &   0.0404(20) &  86.8(2.3) \\
2008 Jun 12 & $\cdots$ &   $-$0.0447(24) &  +0.0082(34) &   0.0454(24) &  84.8(2.1) \\
2008 Jun 13 & $\cdots$ &   $-$0.0434(18) &  +0.0042(30) &   0.0436(18) &  87.2(1.9) \\
2008 Jun 14 & $\cdots$ &   $-$0.0419(23) &  +0.0084(35) &   0.0427(24) &  84.3(2.3) \\
Overall & $\cdots$ & $-$0.04547(87) &  +0.0083(13) & 0.04622(89) & 84.82(80) \\
\hline
2008 Jun 09 & HD 175541 & $-$0.1040(25) & +0.0425(23) & 0.1124(25) & 78.89(60) \\
2008 Jun 10 & $\cdots$ & $-$0.1045(26) & +0.0438(28) & 0.1133(26) & 78.63(71) \\
2008 Jun 11 & $\cdots$ & $-$0.1027(28) & +0.0494(23) & 0.1140(27) & 77.15(61) \\
2008 Jun 12 & $\cdots$ & $-$0.1009(25) & +0.0453(21) & 0.1106(24) & 77.91(56) \\
2008 Jun 13 & $\cdots$ & $-$0.0990(21) & +0.0485(28) & 0.1103(23) & 76.94(71) \\
2008 Jun 14 & $\cdots$ & $-$0.0973(27) & +0.0485(28) & 0.1087(27) & 76.76(72) \\
Overall & $\cdots$ & $-$0.1041(11) & +0.0478(11) & 0.1146(11) & 77.66(26) \\
\hline
2008 Jun 08 & HD 189733 &  $-$0.0174(42) &  +0.0190(31) &  0.0258(37) &  66.3(4.2) \\
2008 Jun 09 & $\cdots$ &  $-$0.0255(12) &  +0.0172(17) &  0.0307(14) &  73.0(1.4) \\
2008 Jun 10 & $\cdots$ &  $-$0.0206(15) &  +0.0125(18) &  0.0241(16) &  74.4(2.1) \\
2008 Jun 11 & $\cdots$ &  $-$0.0209(17) &  +0.0176(17) &  0.0274(17) &  70.0(1.8) \\
2008 Jun 12 & $\cdots$ &  $-$ &  $-$ &  $-$ & $-$ \\
2008 Jun 13 & $\cdots$ &  $-$0.0242(12) &  +0.0163(16) &  0.0292(14) &  73.0(1.5) \\
2008 Jun 14 & $\cdots$ &  $-$0.0209(13) &  +0.0140(13) &  0.0251(13) &  73.1(1.5) \\
Overall & $\cdots$ & $-$0.02476(61) & +0.01706(71) & 0.03007(64) & 72.72(64) \\
\hline
2008 Jun 08 & Cygnus X-1 & +0.6003(51) & $-$4.8315(48) & 4.8687(48) & 138.541(30) \\
2008 Jun 09 & $\cdots$ & +0.7721(34) & $-$5.0076(44) & 5.0668(43) & 139.382(20) \\
2008 Jun 10 & $\cdots$ & +0.5467(40) & $-$4.9797(29) & 5.0096(29) & 138.132(23) \\
2008 Jun 11 & $\cdots$ & +0.6299(40) & $-$4.9701(43) & 5.0099(43) & 138.612(23) \\
2008 Jun 12 & $\cdots$ & +0.6590(40) & $-$5.0088(32) & 5.0519(33) & 138.747(23) \\
2008 Jun 13 & $\cdots$ & +0.5537(43) & $-$4.9368(46) & 4.9677(46) & 138.200(25) \\
2008 Jun 14 & $\cdots$ & +0.6912(39) & $-$4.9267(36) & 4.9749(36) & 138.993(22) \\
Overall & $\cdots$ & +0.6360(51) & $-$4.9579(34) & 4.9985(35) & 138.655(29) \\
\enddata
\label{table2}
\tablecomments{TP is measured by APDs (wavelength range of about 400 to 850 nm), and remaining targets are observed by PMTs (wavelength range of about 400 to 675 nm). TP is subtracted for these targets. Uncertainty in position angle is purely statistical, as position angle is not calibrated absolutely.}
\end{deluxetable}

To determine the unbiased significance of night-to-night variability of each target, I perform a Kolmogorov-Smirnov (K-S) test on the measurement distribution for each pair of nights. This test estimates the significance of variability without prior knowledge of its physical cause. This is because analysis of data from all pairs of nights samples all temporal frequencies available as opposed to focusing on the frequency of an expected signal. For instance, Equations \ref{eqa} through 4 indicate that polarization due to stellar flux scattered off an exoplanetary atmosphere will be strongest at quarter phases and will be zero at conjunctions. Thus, the unbiased K-S test will underestimate the significance of variability caused by this process. However, this test is important in separating stochastic variability of the host star from exoplanetary modulation. \\

\begin{figure}[p]     
 \begin{center}
  \scalebox{0.6}{\includegraphics{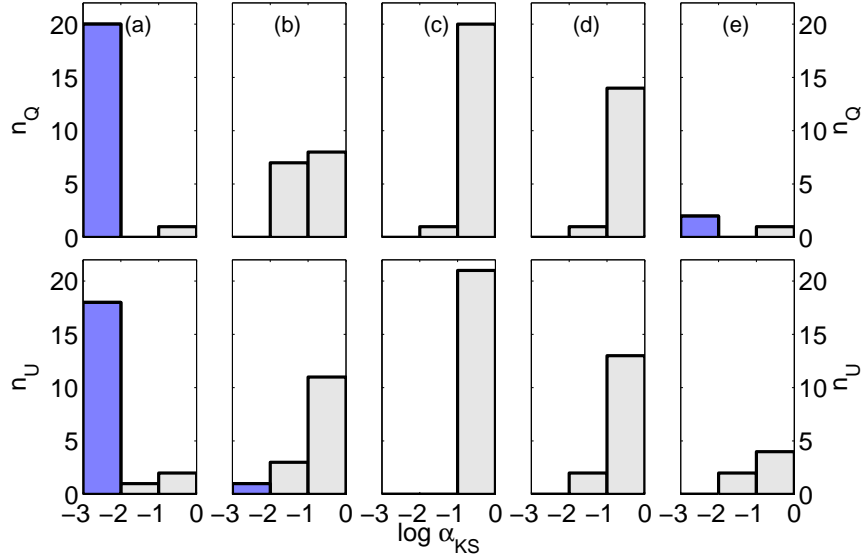}}
  \end{center}
 \caption{Polarimetric variability of observed targets: (a) Cygnus X-1, (b) HD 189733, (c) HD 149026, (d) HD 175541, and (e) HR 8974. Stars with most $\alpha_{\rm KS}$ values less than 0.01 (blue regions) are considered variable. It can be seen that only (a) Cygnus X-1, the variable control source, is significantly variable.}   
 \label{3figs}        
\end{figure}

To claim variability from this test, I require a variability confidence level of $> 99\%$. Therefore, rejection of the null hypothesis with significance $\alpha_{\rm KS} < 0.01$ is required to claim variability of the source. This method is preferred over a $\chi^2$ analysis with a constant polarization model, because no assumptions are made about intra-night measurement distributions and number of measurements (Clarke \& Naghizadeh-Khouei 1994). However, K-S tests of my data show that such measurement distribution is consistent with a Gaussian nature for all targets and for all nights. That is, neither systematic effects nor intra-night variability are significant for any targets. Plotted in Figure \ref{3figs} are the $\alpha_{\rm KS}$ values for all pairs of nights from each star. The likelihood of variability decreases from Figures \ref{3figs}a through \ref{3figs}e. \\

\subsection{HD 189733 (V452 Vul)}

\begin{figure}[p]     
 \begin{center}
  \scalebox{0.55}{\includegraphics{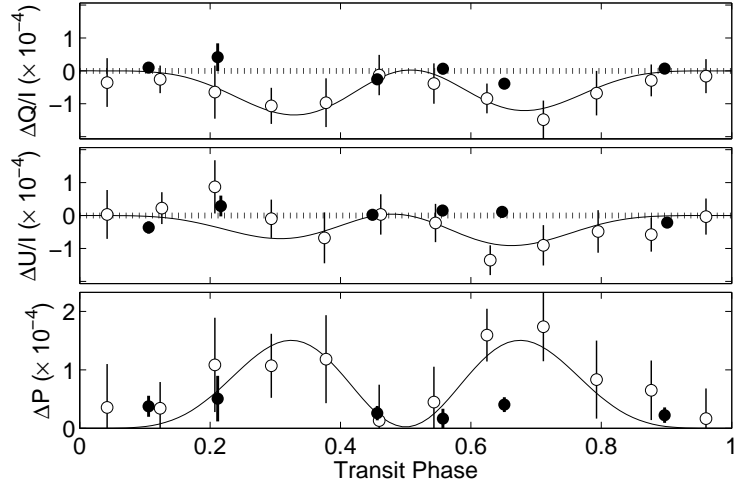}}
  \end{center}
 \caption{Comparison of observed polarization of the HD 189733 transiting hot Jupiter system. Solid circles are data from this work, open circles are data from B08, and the curve represents the exoplanet model of B08. Mean polarizations from each data set have been independently subtracted to show residual polarimetric variability. Phase 0 indicates mid-transit in order to be consistent with B08.}   
 \label{3figq}        
\end{figure}           

\begin{figure}[p]     
 \begin{center}
  \scalebox{0.55}{\includegraphics{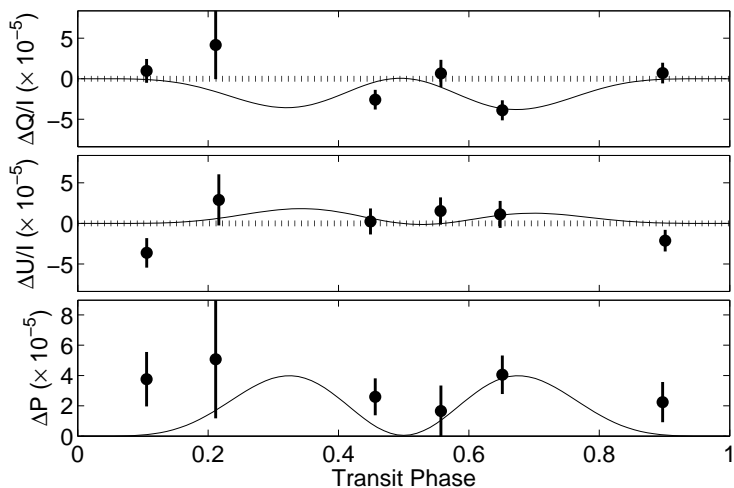}}
  \end{center}
 \caption{Observed polarization of the HD 189733 transiting hot Jupiter system with best fit exoplanet model. Telescope polarization and mean polarization, both effects being at the level of three parts in $10^4$, have been subtracted. Phase 0 indicates mid-transit (inferior conjunction of the exoplanet) in order to be consistent with B08.}   
 \label{3fign}        
\end{figure}           

\begin{figure}[p]     
 \begin{center}
  \scalebox{0.55}{\includegraphics{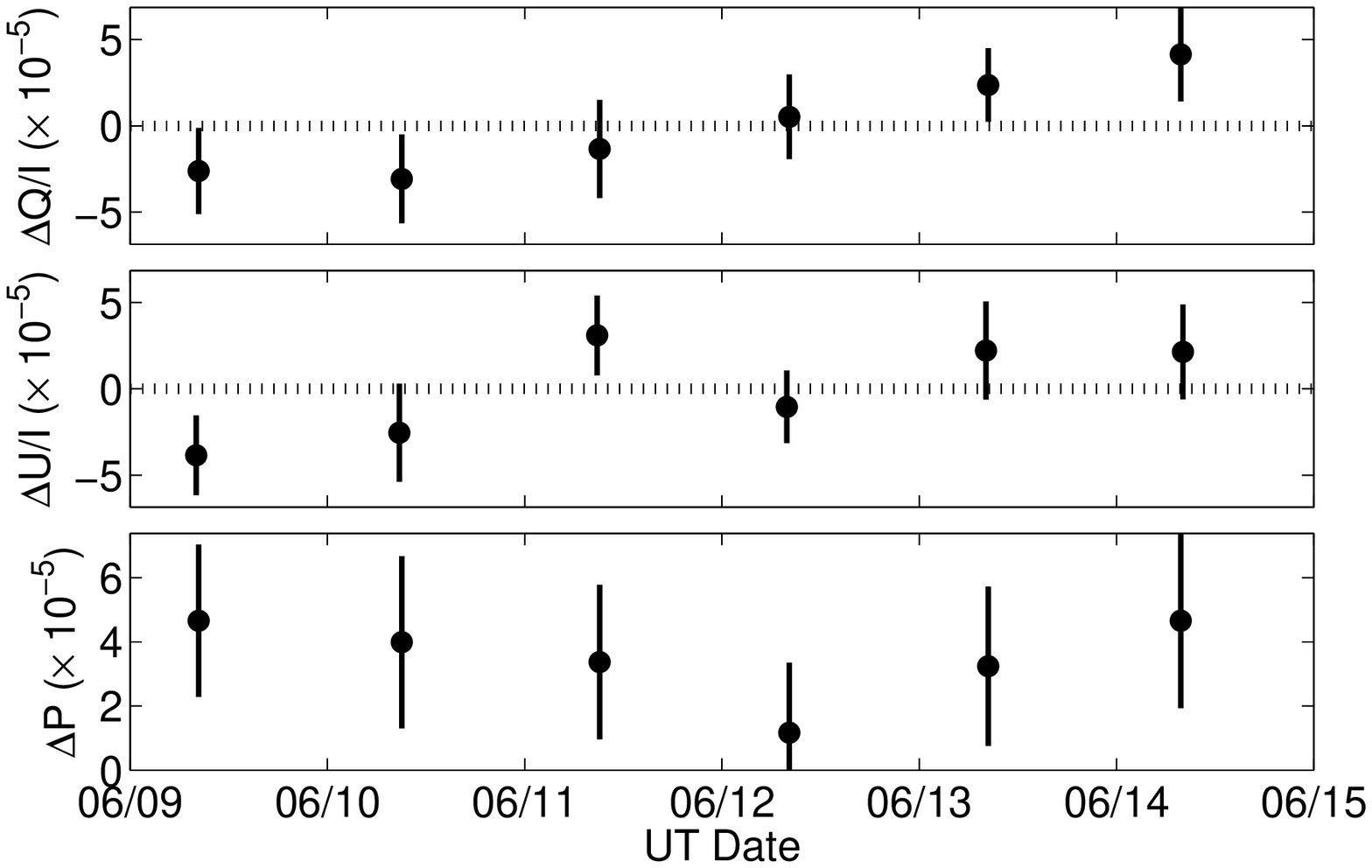}}
  \end{center}
 \caption{Observed polarization of the long period HD 175541 exoplanet system. Telescope polarization as well as residual mean polarization of about 0.1\% have been subtracted.}   
 \label{3figk}        
\end{figure}           

\begin{figure}[p]     
 \begin{center}
  \scalebox{0.65}{\includegraphics[trim = 0in 0.75in 0in 0.35in, clip]{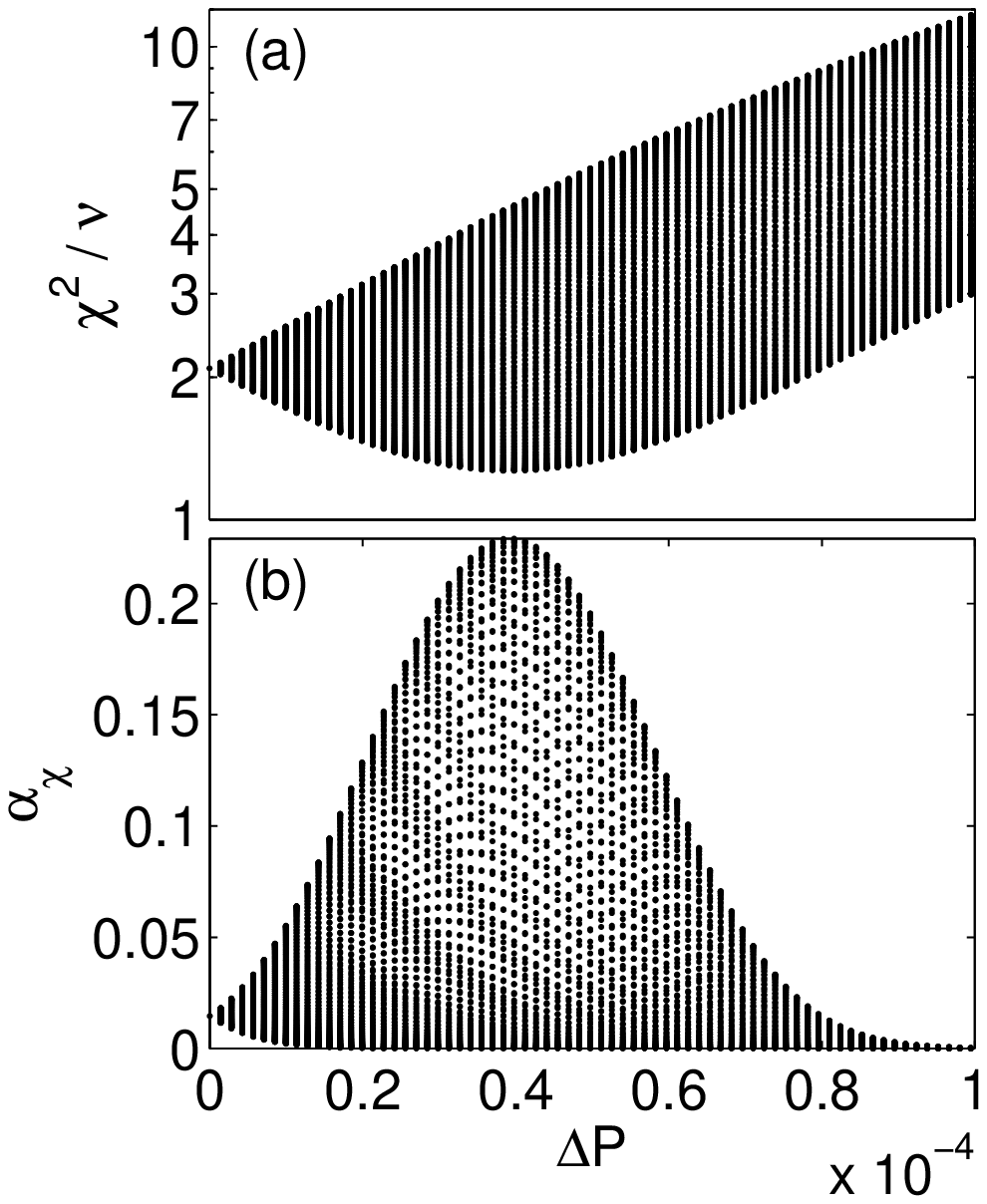}}
  \end{center}
 \caption{Distribution of (a) $\chi^2 / \nu$ and (b) $\alpha_\chi$ as a function of $\Delta P$ for Monte Carlo simulations.}   
 \label{3figu}        
\end{figure}

From Figure \ref{3figs}b, it can be seen that significant variability of the system is not observed. In Figure \ref{3figq}, I show nightly mean polarization of HD 189733 observed with POLISH and compared with the data from B08. It should be noted that the bandpasses are different between B08 ($B$ band) and this work (400 to 675 nm), but the discrepancy in Figure \ref{3figq} cannot simply be due to this. A $\chi^2$ analysis shows that the model reported by B08 to fit their observations fails to accurately reproduce my observations to the $> 99.99\%$ confidence level ($\chi^2 = 99.9$, $\nu = 12$). Figure \ref{3fign} isolates the POLISH data, and Figure \ref{3figk} illustrates the non-variable control system HD 175541 for comparison. To find an upper limit to the polarization amplitude of HD 189733b, I perform Monte Carlo simulations for expected exoplanetary polarization using Equations \ref{eqa} through 4. I set $i = 85 \, \fdg58$ (Torres et al. 2008 analysis of Knutson et al. 2007), $0 \leq \Delta P \leq 3 \times 10^{-4}$, and $0^\circ \leq \Omega \leq 180^\circ$. There exists a reduced $\chi^2$ minimum of $\chi^2 / \nu = 1.27$ with a $\alpha_\chi = 0.229$ probability of a successful fit for $\Delta P = 4.0 \times 10^{-5}$ and $\Omega = 169^\circ$ (Figures \ref{3fign} and \ref{3figu}). This amplitude is larger than the maximum $\Delta P \approx 2 \times 10^{-5}$ allowed for $p = 2/3$ (section 2), indicating that it cannot be due to the planet. Indeed, a constant fit to the data in Figure \ref{3fign} produces a similar reduced $\chi^2$ of 1.79, reinforcing the results of the K-S test that my observations of HD 189733 are consistent with noise. \\

\begin{figure}[p]     
 \begin{center}
  \scalebox{0.5}{\includegraphics[trim = 0in 0.05in 0in 0.35in, clip]{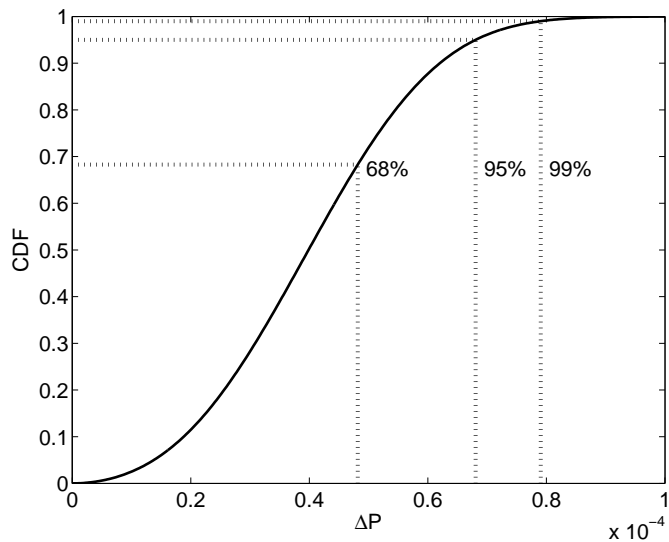}}
  \end{center}
 \caption{Cumulative distribution function for Monte Carlo simulations with polarization amplitude $\Delta P$. Confidence intervals and 68\%, 95\%, and 99\% upper limits on $\Delta P$ are shown.}   
 \label{3figt}        
\end{figure}

Upper limits to the polarization amplitude can be determined from the cumulative distribution function of $\alpha_\chi$ as a function of $\Delta P$ (Figure \ref{3figu}b), which is shown in Figure \ref{3figt}. The 68\%, 95\%, and 99\% upper limits to exoplanetary polarimetric modulation are $\Delta P < (4.8$, $6.8$, and $7.9) \times 10^{-5}$, respectively. Such large upper limits exist because I was unable to observe the system where exoplanetary polarization is maximized, which is near both quarter phases. Indeed, the upper limits are based significantly on the single night of observations at phase 0.65, which is near last quarter (Figure \ref{3fign}). Since stellar activity inhibits albedo constraint from photometry (Rowe et al. 2008a), future observations of this system with POLISH are necessary in order to provide an albedo constraint. However, my observations have ruled out the $\Delta P \approx 2 \times 10^{-4}$ amplitude of modulation reported by B08 with $> 99.99\%$ confidence. \\

A potential cause of polarimetric variability in hot Jupiter host stars is starspot activity. Photometric observations with the MOST satellite suggest the existence of starspots on the short period $\tau$ Bo\"{o} that follow the rotation period of the star (Walker et al. 2008). There is also some evidence that Ca II H and K emission from the short period HD 179949 may follow the stellar rotation period (Shkolnik et al. 2005, 2008). HD 187933 itself is known to be active, with up to $1\%$ of its surface covered in spots at any time (H\'{e}brard \& Lecavelier des Etangs 2006; Croll et al. 2007; Pont et al. 2007; Winn et al. 2007; Moutou et al. 2008). These spots appear to rotate with the roughly 11.8 day stellar rotation period (Henry \& Winn 2008; Croll et al. 2008). The run-averaged polarization of HD 189733, observed by both B08 as well as POLISH, is an order of magnitude larger than those seen in stars at comparable distances (Hough et al. 2006; Lucas et al. 2009). This enhanced polarization may be due to starspots. \\

Since optical depth of a stellar atmosphere decreases from center to limb, and since the scattering angle increases to $90^\circ$ at the limb, stellar polarization is dominated by the limb (Carciofi \& Magalh\~{a}es 2005, hereafter CM05). Spatially resolved polarimetry is possible for the Sun, and this effect has been observed (Faurobert et al. 2001; Faurobert \& Arnaud 2003). Just after a starspot appears above the limb and just before it disappears below the limb, the greatest asymmetry in limb polarization occurs. Indeed, CM05 model the polarized signal from an exoplanetary transit to be strongest at ingress and egress, when part of the limb is occulted. The radial polarization profile of most stars is unknown, but the models of CM05 indicate that an exoplanet-sized starspot on HD 189733 may generate a polarimetric amplitude of $10^{-4}$ to $10^{-5}$ when near the stellar limb. The unknown stellar latitude and stability of these spots must generate complex variations in stellar polarimetry. While starspots on HD 189733 appear to be tied to the stellar rotation period, their effects may not average out even during long observing campaigns. Unfortunately, B08 do not discuss the possibility of starspots causing their observed modulation. \\

\subsection{HDE 226868/Cygnus X-1 (V1357 Cyg)}

I verify the exceedingly large significance of polarimetric variability of this system in both Stokes parameters (Figure \ref{3figs}a) as well as its roughly phase-locked amplitude of variability of $\Delta P \approx 0.1\%$ (Table \ref{table2}). Further analysis of the modulation of this system, including inclination estimation, will be left for a forthcoming paper. \\

\subsection{HD 149026}

A short period, transiting exoplanet orbits this weakly polarized star (Sato et al. 2005). Significant variability of this system is not observed to a precision of $\Delta P_{\rm obs} = 1.6 \times 10^{-5}$ (Figure \ref{3figs}c and Table \ref{table3}). Unfortunately, my lack of observations near quarter phase inhibits constraint of the exoplanetary albedo. \\

\subsection{HD 175541}

This weakly polarized star harbors a long-period exoplanet (Johnson et al. 2007). Since the expected polarimetric signal due to the exoplanet is $\Delta P_{\rm exp} \la 2 \times 10^{-8}$ over an entire orbit, any observed polarimetric variability from the system cannot be due to the exoplanet. This conclusion is bolstered by the lack of significant variability of the system: I observe no significant variability above the level of $\Delta P_{\rm obs} = 2.6 \times 10^{-5}$ (Figure \ref{3figs}d and Table \ref{table3}). The trend seen in Stokes $Q$ lies at low significance (Figures \ref{3figs}d and \ref{3figk}). \\

\subsection{HR 8974 ($\gamma$ Cep, HD 222404)}

Hatzes et al. (2003) discovered an exoplanetary companion to the primary component of this unpolarized binary system. I expect the amplitude of the exoplanetary polarimetric signal over an entire orbit to be $\Delta P_{\rm exp} \la 5 \times 10^{-9}$ and consequently undetectable. Indeed, examination of Figure \ref{3figs}e and Table \ref{table3} shows that this system is not significantly variable to a precision of $\Delta P_{\rm obs} = 9.0 \times 10^{-6}$. \\
 
\section{Conclusion}

I do not detect the $\Delta P \approx 2 \times 10^{-4}$ polarimetric modulation of the hot Jupiter system HD 189733 reported by B08 with $> 99.99\%$ confidence. Using the high precision polarimeter POLISH, I derive an upper limit to the polarimetric modulation of the exoplanet of $\Delta P < 7.9\times 10^{-5}$ with 99\% confidence. I am unable to constrain the albedo of the exoplanet; future observations of the system polarization near quarter phases (phases 0.25 and 0.75) are required. \\

The amplitude of the signal reported by B08 is at least one order of magnitude larger than expected given upper limits to the albedos of three other hot Jupiters (Leigh et al. 2003a, 2003b; Rowe et al. 2008b; Lucas et al. 2009). In addition, my significant non-detection shows the modulation reported by this group cannot be caused by the exoplanet. Starspot activity on HD 189733 has the potential to introduce polarimetric modulation with an amplitude of one part in $10^4$ to $10^5$ (cf. CM05). While starspots on HD 189733 appear to share the stellar rotation period, their unknown latitudinal distribution and timescale of stability suggest that their effects may not simply average out over long observing campaigns. However, my data are acceptably explained by a constant polarization model, so polarimetric variability of the HD 189733 system is not statistically significant to the $\Delta P_{\rm obs} = 2.1 \times 10^{-5}$ level. \\

My observations of the transiting hot Saturn HD 149026b encompass neither first nor last quarter phases, so I cannot constrain the geometric albedo of this exoplanet. I have observed no significant, polarimetric variability of the short period exoplanets HD 189733b and HD 149026b nor of the long period exoplanets HD 175541b and HR 8974b. However, polarimetric variability of the variable control source Cygnus X-1 is detected with high confidence. Gaussian distribution of nightly measurements shows that no observed target, even the famously variable Cygnus X-1, exhibits significant polarimetric variability during a single night. \\

High precision, polarimetric monitoring of both transiting and non-transiting, short and long period exoplanets is a new field that is poised for major discoveries. The first detection of exoplanetary optical, scattered light has been made, and it will soon be detected with polarimetry. The atmospheres and accurate masses of close-in exoplanets will be probed. The scattering properties of exoplanet host star atmospheres will be studied from polarized transit observations, which will be valuable to calibrate high precision, photometric observations. The atmosphere of the Earth is effectively removed by high precision polarimetry; therefore, such observations allow ground-based observatories to make important contributions to exoplanetary science that complement the ever-increasing stream of discoveries with space-based observatories. \\

 \acknowledgements
 I would like to thank Shrinivas Kulkarni and James Graham for their support as well as Bruce Macintosh and Lynne Hillenbrand for suggestions on this manuscript. I would also like to acknowledge funding from the Moore Foundation as well as from the Center for Adaptive Optics. This research has made use of the SIMBAD database, operated at CDS, Strasbourg, France. \\
 
 {\it Facilities:} \facility{Hale}.

 \end{document}